# Dust as the cause of spots on Jupiter

G.B. Field[1], G. P. Tozzi[2], and R. M. Stanga[3]

[1] Harvard - Smithsonian Center for Astrophysics, 60 Garden St., Cambridge, MA 02138 and Osservatorio Astrofisico di Arcetri
[2] Osservatorio Astrofisico di Arcetri, Largo Enrico Fermi 5, I - 50125, Firenze (I)
[3] Dipartimento di Astronomia dell'Università - Firenze (I)

**Abstract.** The long-lived spots caused by the impact of fragments of Comet S-L 9 on Jupiter can be understood if clouds of dust are produced by the impact. These clouds reside in the stratosphere, where they absorb visible light that would ordinarily reflect from the cloud deck below, and reflect radiation at infrared wavelengths that would ordinarily be absorbed by atmospheric methane. Here we show that, provided that the nucleus of a fragment is composed substantially of silicates and has a diameter greater than about 0.4 km, dust in the required amounts will condense from the hot gas composed of cometary and Jovian material ejected from the site where the fragment entered, and the dust will be suspended in the stratosphere for long periods. Particles about 1 $\mu$m in radius can explain both the optical properties and longevities of the spots. According to our model, a silicate band should be present in the $10 - \mu$m spectra of the spots.

## 1. Introduction

Observations reported to the University of Maryland S-L 9 Network show that after a brief period following impact, spots form that persist for many Jupiter rotation periods, that are brighter than the cloud layer in the surrounding atmosphere at wavelengths in the methane band, and are darker than the clouds in the visible band. It is reasonable to interpret these spots as due to stratospheric dust because dust can reflect and absorb sunlight as required, it can be produced by condensation as hot gas ejected by the impact cools, and it will remain in the stratosphere for long periods of time.

The source of dust could in principle be the coma of the fragment, composed of particles so small that they survive entry into the upper atmosphere of Jupiter, as do micrometeorites on Earth. However, the photometric properties of the coma of the fragments before impact show that it is too small by a factor of $10^6$ to explain the data. Alternatively, the dust may have condensed from vapor contained in the hot gas that is ejected (Ahrens et al. 1994; Takata et al. 1994; Mac Low & Zahnle 1994; Zahnle & Mac Low 1994) and spread over a large area of the upper atmosphere following the entry of each fragment. This gas contains both atmospheric material and material from the vaporized fragment. Given the presence of uncondensed $CH_4$ and clouds of $NH_3$, $NH_4SH$, and $H_2O$ at various depths in the atmosphere (Chamberlain & Hunten 1987) the atmospheric material will certainly contain atoms of C, and, depending on the depth of penetration of the fragment, N, S, and O in addition to $H_2$ and He. The vaporized fragment will contain the constituents of silicates (Si, Mg, Fe, and O) as well as additional C, N, and O from any icy component of the fragment (Festou et al. 1994). As the vapor containing these atoms cools from a high (several $10^3$K) temperature, molecules of $Mg_2SiO_4$, FeS, $H_2O$, $NH_3$, and carbon compounds will form. Observations by HST and IUE have demonstrated the presence of Mg, normally a component of silicates (IAU Circ. # 6027, 6038, and 6041), so silicates may form and condense as dust particles in the stratosphere.

In this paper we consider only the contribution of material in the fragment to the dust. For simplicity, we will focus on one of the chief constituents of cometary material, silicates such as $Mg_2SiO_4$. We will assume that the total mass $M$ of the nucleus is vaporized and ejected to form a layer of radius $R$ and area $A = \pi R^2$. After condensation into particles of radius $a$ has occurred, the mean optical depth of the layer at a wavelength $\lambda$ will be

$$\bar{\tau} = \frac{1}{A}\int \tau dA = \frac{\sigma}{Am_d}\int \rho dA dz = \frac{\sigma M}{Am_d}, \qquad (1)$$

where

$$\sigma = \pi a^2 Q \qquad (2)$$

is the appropriate cross section of the dust grain at wavelength $\lambda$, $Q$ is the efficiency factor (Spitzer 1978)

$$m_d = \frac{4}{3}\pi a^3 \rho_i \qquad (3)$$

is the mass of a spherical grain of radius $a$, $\rho_i = 3$ g cm$^{-3}$ is the density of silicate, and $\rho$ is the density of silicate vapor in the hot gas. It follows that the mass of a fragment required to explain a spot of mean optical depth $\bar{\tau}$ is

$$M = \frac{m_d}{\sigma}\bar{\tau}A = \frac{4\rho_i a \bar{\tau} \pi R^2}{3Q} \approx 10^{19}\frac{R_4^2 \bar{\tau} a}{Q} \text{ g}, \qquad (4)$$

where $R_4$ is $R/10^4$ km.

Further progress requires knowledge of $\bar{\tau}$, $a$, and $Q$. We reason as follows. Dust does not reflect or absorb readily unless $2\pi a/\lambda \geq 1$ (Spitzer 1978). Since the spots are conspicuous in the methane band at $\lambda = 2.4$ $\mu$m, we may assume that $a \geq 0.4$ $\mu$m. Such a particle size guarantees that $2\pi a/\lambda$ is also substantial at visible wavelengths, as required by the observations (Noll 1994), so we may take $Q_V \simeq 1$ at those wavelengths (Spitzer 1978). Since the spots are conspicuous but semitransparent at visible wavelengths, in the absence of quantitative data we estimate very roughly that $\bar{\tau}_V \simeq 0.3$. It follows from equation (4) that if $R_4 \sim 1$,

$$M > 1.2 \times 10^{14} \text{ g}, \qquad (5)$$

so the diameter of a typical fragment then satisfies

$$D = \left(\frac{6M}{\pi \rho_i}\right)^{1/3} > 0.4 \text{ km}, \qquad (6)$$

a not unreasonable value (Sekanina et al. 1994). We note that if $a < 2$ $\mu$m, as we find below to explain the longevity of the spots, $2\pi a/\lambda$ can still be large at wavelengths up to 12 $\mu$m. We also note that the value of $D$ indicated by equation (6) is a lower limit for an additional reason: it would be still larger if not all the material in the fragment is vaporized, ejected, and condensed into observable particles as we have assumed (see below).

Can the grains grow to micron size in the time available? To answer this, we assume that after spreading laterally, the gas cools through 1400 K, the condensation temperature of silicates (Field 1974), contracting vertically in hydrostatic equilibrium as it does so. When the temperature $T$ falls below 1400 K, the vapor becomes supersaturated, spontaneous nucleation occurs, and dust grains grow (Draine 1979) at a rate that depends only upon $T$ and the dimensionless time variable,

$$\eta = \frac{1}{3}\pi r^2 nut, \qquad (7)$$

where $r = (3\mu_c m_H/4\pi\rho_i)^{1/3} = 1.5 \times 10^{-8}$ cm is the radius of a lattice site in terms of the molecular weight $\mu_c$ of the constituents of the grain (about 26 if the latter are mainly Mg and Si), $n = \rho/\mu_c m_H$ is the density of atoms in the vapor, $u = (8RT/\pi\mu_c)^{1/2} = 3 \times 10^3 T^{1/2}$ cm sec$^{-1}$ is the mean speed of atoms in the vapor, and $t$ is the time. An approximate solution to the equation governing the growth of the mean grain radius is (Draine 1979)

$$a(\eta) = 0.23\left(\frac{\theta}{T}\right)\left(\frac{\eta}{\ln\eta}\right)\left[1 - e^{-\alpha(\eta)}\right]^{1/3} r, \qquad (8)$$

where $\alpha(\eta)$ is large in our case, so we neglect $e^{-\alpha}$ in what follows. Here $\theta$ is 2.7 times the free energy (in units of the Boltzmann constant) of each lattice site on the surface.

We can use equation (8) to find the time $t$ for $a(t)$ to reach the value $a$ required by the observations. The column density of vapor averaged over a spot can be written

$$\langle \int \rho dz \rangle = m_H \mu_c \langle \int n dz \rangle \simeq m_H \mu_c H \bar{n}, \qquad (9)$$

where $\bar{n}$ is the mean density of atoms of vapor, $H = RT/\mu_p g = 7 \times 10^3 T$ cm is the scale height of the vapor layer, and $\mu_p \simeq 4$ is the molecular weight of a somewhat arbitrary equal mixture by mass of H$_2$ and silicate vapor. Equation (9) allows us to write equation (1) in the form

$$\bar{\tau} = \frac{3Q m_H \mu_c H \bar{n}}{4\rho_i a}. \qquad (10)$$

If we substitute $\bar{n}$ from equation (10) into equations (7) and (8) we can write

$$\frac{a(t)}{a} = \frac{0.23(\theta/T)\bar{\tau} u t}{3Q \ln\eta H}, \qquad (11)$$

where we have used $4\pi r^3 \rho_i/3 = m_H \mu_c$. At $T = 1400$ K, the condensation temperature for silicates, we have $u = 1.2 \times 10^5$ cm s$^{-1}$ and $H = 10^7$ cm. If we assume that $a \simeq 1$ $\mu$m (see above), we can iterate equation (8) with $\theta/T \simeq 10$ (Draine 1979) to get $\ln\eta = 10$. Taking $Q \simeq 1$ (see above), we then find that

$$\frac{a(t)}{a} \simeq 9 \times 10^{-4} \bar{\tau} t, \qquad (12)$$

so that the time for $a(t)$ to reach its final value is

$$t \sim \frac{18}{\bar{\tau}} \text{ min}. \qquad (13)$$

For $\bar{\tau} \sim 0.3$, this time is about an order of magnitude larger than is indicated by the fact that spots appeared a few minutes after impact. This discrepancy cannot be explained by postulating a smaller size for the particles observed in the infrared, as explained above, but it might be explained if some of the vapor condensed into particles that are so small that they do not contribute in the infrared. This would allow the density of vapor to be larger than we have estimated from the observed optical depth, and hence the time required to reach the prescribed value of $\eta$ according to equation (7) to be smaller. Note that this interpretation would strengthen the lower limit on $D$ in equation (6).

The dust particles drift downward under gravity until they reach the tropopause, which is located $\sim$ 20 km above the ammonia clouds (Chamberlain & Hunten 1987), after which they are dispersed by convective motions in the troposphere. We can learn something about the particles by requiring that the time for removal from the stratosphere $t_R$ is at least as long as 50 days, a lower limit on the observed lifetime of the larger spots (Noll 1994). The drag force preventing the particles from falling depends on whether the local mean free path of molecules in the atmosphere, $mfp$, is larger or smaller than the particle radius $a$. We find that the $mfp \sim 0.5$ $\mu$m at Jupiter's tropopause, and that it increases like $e^{z/H}$ above that level, where $H = RT/\mu g = 20$ km is the scale height of the stratosphere ($\mu = 2.3$ and mean $T = 150$ K according to Chamberlain & Hunten (1987)). Suppose that $a < 0.5$ $\mu$m. As this implies that $a < mfp$ at all levels of the stratosphere, the applicable formula for the drag force is then (Spitzer 1978)

$$F = \frac{4}{3}\pi \rho a^2 uv, \qquad (14)$$

where $v$ is the drift speed.

Setting this equal to the gravitational force $m_d g = (4/3)\pi \rho_i a^3 g$, we get

$$v = \frac{\rho_i g a}{\rho u}, \qquad (15)$$

where $\rho_i = 3$ g cm$^{-3}$ is the internal density of a silicate particle. If a particle starts at altitude $z$, the time to fall to the tropopause is

$$t_F = \frac{H \rho_0 u}{a \rho_i g}\left(1 - e^{-z/H}\right), \qquad (16)$$

where $\rho_0 = 3.7 \times 10^{-5}$ g cm$^{-3}$ is the atmospheric density at the tropopause (Chamberlain and Hunten 1987). With $T = 150$K and $\mu = 2.3$, $u = (8RT/\pi\mu)^{1/2} = 1.2 \times 10^5$cm s$^{-1}$, and we find

$$t_F = \frac{9 \times 10^6}{a_\mu}\left(1 - e^{-z/H}\right) \leq \frac{9 \times 10^6}{a_\mu}\text{s}, \qquad (17)$$

where $a_\mu = (a/1\ \mu\text{m})$. Particles larger than 0.5 $\mu$m fall at the same rate until they reach a level where $mfp \sim a$, after which according to Stokes's law they fall faster than implied by equation (15). It follows that equation (17) gives an upper limit on $t_F$ for all values of $a$. Since other processes, such as convection that penetrates the stratosphere, could remove particles in even shorter times than $t_F$, $t_F$ is also an upper limit on $t_R$, the removal time.

Since the latter is at least 50 days = $4 \times 10^6$s (Noll 1994), we have

$$4 \times 10^6 \text{s} \leq t_R \leq t_F \leq \frac{9 \times 10^6}{a_\mu}\ \text{s}, \qquad (18)$$

so

$$a \leq 2\ \mu\text{m}. \qquad (19)$$

If this inequality were violated, particles would be removed in less than 50 days, whatever their initial altitude and whatever processes may speed up their fallout.

If we put $a < 2$ $\mu$m and $Q \sim 1$ in equation (4), we obtain an upper limit on the mass of silicate dust required to explain the observations, namely, $2 \times 10^{15} R_4^2 \bar{\tau}$ g or $\sim 6 \times 10^{14}$g if $R_4 \sim 1$ and $\bar{\tau} \sim 0.3$. Note, however, that this value cannot be used as an upper limit to the mass of the fragment because of the possible inefficiencies in producing observable particles referred to above.

We conclude that if the spots are caused by dust condensed from cometary material, the fragment that caused each spot must have a diameter that exceeds 0.4 km. If, as we have assumed, the fragment was composed at least partly of silicates, a silicate feature should be present in 10 - $\mu$m spectra[1]. Finally, observations of the longevity of spots, together with their optical properties, constrain the sizes of the particles responsible for the infrared spots to be of the order of one micron.

*Acknowledgements.* We acknowledge stimulating conversations with Susan Field, Harold Weaver, Keith Noll, Michael A'Hearn, Andrea Ferrara, and George Rybicki. This research was supported by NASA.

## References

Ahrens, T. J., Takata, T., O'Keefe, J. D., Orton, G. S., 1994, Geophys. Res. Lett. 21, 1087
Chamberlain, J. W., Hunten, D. M., 1987, Theory of Planetary Atmospheres. Academic Press, San Diego, p. 49
Draine, B. T., 1979, Ap&SS 65, 313
Festou, M. C., Rickman, H., West, R. M., 1994, A&AR 4, 363
Field, G. B., 1974, ApJ 187, 453
Mac Low, M. - M., Zahnle, K., 1994, ApJ Lett., in press
Nicholson, P. D., 1994, personal communication
Noll, K., 1994, personal communication
Sekanina, Z., Chodas, P. W., Yeomans, D. K., 1994, AJ, in press
Spitzer, L. Jr. 1978, Physical Processes in the Interstellar Medium. Wiley, New York
Takata, T., O'Keefe, J. D., Ahrens, T. J., Orton, G. S., 1994, Icarus, in press
Zahnle, K., Mac Low, M. - M., 1994, Icarus 108, 1

This article was processed by the author using Springer-Verlag LaTeX A&A style file *L-AA* version 3.

---

[1] Since this paper was submitted, Nicholson (1994) has reported that the Palomar group has found evidence of the silicate band in spectra of the impact of fragment R.